\documentclass[a4paper,twoside]{article}
\newif\ifdraft \global\drafttrue
\def\production{\global\draftfalse}
\production
\usepackage[T1]{fontenc}
\usepackage[latin1]{inputenc}
\usepackage{a4wide}
\usepackage{times}
\usepackage{graphicx}
\usepackage{theorem}
\usepackage[colorlinks=true,hyperindex=true]{hyperref}
\usepackage{color}
\usepackage{fancyhdr}
\usepackage{latexsym}
\usepackage{amsmath}
\usepackage{bbm}
\usepackage{amssymb}
\usepackage{amsfonts}
\usepackage{enumerate}
\usepackage{cancel}
\ifdraft
\usepackage{showkeys}
\fi
\newcounter{smallarabics}

\newcounter{smallroman}

\newcommand{\ben}{\begin{enumerate}[{\rm (1)}]}
\newcommand{\een}{\end{enumerate}}

\newtheorem{theoreme}{Theorem }[section]
\newtheorem{proposition}[theoreme]{Proposition}
\newtheorem{lemma}[theoreme]{Lemma}
\newtheorem{definition}[theoreme]{Definition}
\newtheorem{corollary}[theoreme]{Corollary}

\def\rr{{\mathbb R}}
\def\zz{{\mathbb Z}}
\def\cc{{\mathbb C}}

\def\textsl{{}}
\let\oldi\i
\newcommand{\myi}{\oldi}

\def\Im{{\rm Im}\,}

\def\ch{\mathfrak{h}}

\newcommand{\slim}{\mathop{\mathrm{s-lim}}\limits}

\def\c0inf{C_0^\infty}
\def\bep{\begin{proposition}}
\def\eep{\end{proposition}}

\def\cH{{\cal  H}}

\def\i{{\rm i}}
\newcommand{\beq}{\begin{equation}}
\newcommand{\eeq}{\end{equation}}
\newcommand{\bear}[1]{\begin{array}{#1}}
\newcommand{\ear}{\end{array}}

\def\sp{{\hat e}}

\newcommand{\e}{\mathrm{e}}
\renewcommand{\i}{\mathrm{i}}

\renewcommand{\d}{\mathrm{d}}

\def\qed{$\Box$\medskip}

\def\bel{\begin{lemma}}
\def\eel{\end{lemma}}
\def\bet{\begin{theoreme}}
\def\eet{\end{theoreme}}
\def\bed{\begin{definition}}
\def\eed{\end{definition}}
\def\bar{\overline}

\def\12{\frac{1}{2}}

\def\e{{\rm e}}

\def\d{{\rm d}}
\def\Ran{{\rm Ran}\,}

\def\one{{\mathbbm 1}}

\def\cH{{\cal H}}

\def\Ker{{\rm Ker}\,}

\def\ac{{\rm ac}}

\def\sp{{\rm sp}}
\def\cS{{\cal S}}

\def\fh{\mathfrak{h}}

\def\tr{{\rm tr}}

\def\cO{{\cal O}}

\def\ie{{\sl i.e., }}
\def\ES{{\rm ES}}
\def\FCS{{\rm FCS}}
\def\GC{{\rm GC}}

\begin{document}
\def\today{}
\title{Entropic fluctuations in XY   chains\\and reflectionless Jacobi matrices}
\author{V. Jak\v{s}i\'c$^{1}$,  B. Landon$^{1}$,  C.-A. Pillet$^2$, 
\\ \\ 
$^1$Department of Mathematics and Statistics\\ 
McGill University\\
805 Sherbrooke Street West \\
Montreal,  QC,  H3A 2K6, Canada
\\ \\
$^2$Aix-Marseille Universit\'e, CNRS UMR 7332, CPT, 13288 Marseille, France\\
Universit\'e du Sud Toulon-Var, CNRS UMR 7332, CPT, 83957 La Garde, France
}
\maketitle
{\small
{\bf Abstract.} We study entropic functionals/fluctuations  of the XY chain with Hamiltonian
$$\frac{1}{2}\sum_{x\in \zz}J_x\left(\sigma_x^{(1)}\sigma_{x+1}^{(1)}
+\sigma_x^{(2)}\sigma_{x+1}^{(2)}\right)+\lambda_x\sigma_x^{(3)}
$$
where initially the left $(x\leq 0)$/right $(x>0)$ part of the chain is in thermal equilibrium at
inverse temperature $\beta_l/\beta_r$. The temperature differential results in a non-trivial
energy/entropy flux across the chain. The Evans-Searles (ES) entropic functional describes
fluctuations of the flux observable with respect to the initial state while the Gallavotti-Cohen (GC) 
functional describes these fluctuations with respect to the steady state (NESS) the chain reaches in the
large time limit. We also consider the full counting statistics (FCS) of the energy/entropy flux
associated to  a repeated measurement protocol, the variational entropic functional 
(VAR) that arises as the quantization of the variational characterization of the classical
Evans-Searles functional and a natural class of entropic functionals that interpolate between FCS and
VAR. We compute these functionals in closed form in terms of the scattering data of the Jacobi 
matrix $hu_x=J_xu_{x+1}+\lambda_xu_x +J_{x-1}u_{x-1}$ canonically
associated to the XY chain. We show that all these functionals are identical if and only if $h$ is
reflectionless (we call this phenomenon entropic identity).
If $h$ is not reflectionless, then the ES and GC functionals remain equal but
differ from the FCS, VAR and interpolating functionals. Furthermore, in the non-reflectionless case, 
the ES/GC functional does not vanish at $\alpha=1$ (\ie the Kawasaki identity fails) and does not 
have the celebrated
$\alpha \leftrightarrow 1-\alpha$ symmetry. The FCS, VAR and interpolating functionals always have 
this symmetry. In the Schr\"odinger case, where $J_x=J$ for all $x$, the entropic identity leads to
some unexpected open problems in the spectral theory of one-dimensional discrete Schr\"odinger
operators.} 

\thispagestyle{empty}
\section{Introduction}

The XY chain is an exactly solvable model which  is often used to illustrate  
existing theories and to test emerging theories in quantum statistical mechanics. The literature 
on XY chains is enormous and we mention here the mathematically rigorous works
\cite{Ar3, Ar4, AB1, AB2, AH, AP, BM, HR, Ma, Mc, OM}. It is therefore not surprising that a model
of an open quantum system based on the XY chain was also a testing ground for the recent 
developments in non-equilibrium quantum statistical mechanics initiated in the works
\cite{JP1, JP2, Ru2}. More precisely, the first proofs of the existence of a non-equilibrium steady 
state (NESS) and of strict positivity of entropy production in an open quantum system were given 
in the papers \cite{AH,AP} in the context of XY  chains. The purpose of this paper is similar and 
our goal is to test the emerging theory of entropic fluctuations in non-equilibrium quantum statistical 
mechanics developed in \cite{JOPP} (see also \cite{DDM,Ku,Ro,TM} for related works) on the example 
of XY chains. For discussion and references to the classical theory of entropic fluctuations 
(which, starting with the seminal works \cite{CG,ES}, has played a dominant role in the recent 
theoretical, numerical and experimental advances in classical non-equilibrium statistical mechanics) 
we refer the reader to the reviews \cite{JPR,RM}.

The paper is organized as follows. In Section~\ref{sec-confXY} we describe the XY chain confined 
to a finite interval in $\zz$, introduce the finite volume and finite time entropic functionals and 
describe their basic properties. The thermodynamic limit of the confined XY chain and its entropic 
functionals is discussed in Section~\ref{sec-thermlimit}. The elements of spectral and scattering 
theory needed to state and prove our results are reviewed in Section~\ref{sec-operator}. In 
Section~\ref{sec-ness} we review the results regarding the existence of the NESS and introduce the 
finite time Gallavotti-Cohen functional. Our main results are stated in Section~\ref{sec-timelimit}. 
In Section~\ref{sec-discussion} we briefly discuss these results and in Section~\ref{sec-reflectionless}
we comment on reflectionless Jacobi matrices. The proofs are given in Section~\ref{sec-proofs}. 
Some proofs are only sketched and the reader can find details and additional information in the 
forthcoming Master's thesis \cite{La}. 
\bigskip

{\noindent\bf Acknowledgment.}  
The research of V.J. and B.L. was partly supported by NSERC. The research of C.-A.P. was partly
supported by ANR (grant 09-BLAN-0098). We wish to thank C.~Remling and B.~Simon for useful 
discussions.
\subsection{The confined  XY chain} 
\label{sec-confXY} 

To each $x\in \zz$ we associate the Hilbert space $\cH_x=\cc^2$ and the corresponding matrix algebra
${\cal O}_x={\rm M}_2(\cc)$. The Pauli matrices
$$
\sigma_x^{(1)}=\left[\begin{array}{cc}0&1\\1&0\end{array}\right],\quad
\sigma_x^{(2)}=\left[\begin{array}{cc}0&-\i\\\i&0\end{array}\right],\quad
\sigma_x^{(3)}=\left[\begin{array}{cc}1&0\\0&-1\end{array}\right],
$$
together with the identity matrix $\one_x$ form a basis of ${\cal O}_x$ that satisfies the relations 
$$
\sigma_x^{(j)}\sigma_x^{(k)}=\delta_{jk}\one_x+\i\varepsilon^{jkl}\sigma_x^{(l)}.
$$

Let $\Lambda=[N, M] $ be a finite interval in $\zz$. The Hilbert space and the algebra of observables 
of the XY chain confined to $\Lambda$ are 
\[
\cH_\Lambda=\bigotimes_{x\in\Lambda}\cH_x, \qquad 
{\cal O}_\Lambda=\bigotimes_{x\in\Lambda}{\cal O}_x.
\]
The spectrum of the observable $A\in{\cal O}_\Lambda$ is denoted $\sp(A)$. If $A$ is self-adjoint,
$\one_{E}(A)$ denotes the spectral projection associated to $E\in\sp(A)$.

We shall identify $A_x\in\cO_x$ with the element
$(\otimes_{y\in \Lambda \setminus\{x\}}\one_y)\otimes A_x$ of $\cO_\Lambda$. 
In a similar way we identify ${\cal O}_{\Lambda^\prime}$ with the appropriate subalgebra of 
${\cal O}_\Lambda$ for $\Lambda^\prime \subset \Lambda$.
With these notational conventions the Hamiltonian of  the XY chain confined to $\Lambda$ is 
\begin{equation*}
H_\Lambda=\frac{1}{2}\sum_{x\in[N,M[}J_x\left(
\sigma_x^{(1)}\sigma_{x+1}^{(1)}+\sigma_x^{(2)}\sigma_{x+1}^{(2)}
\right)+\frac12\sum_{x\in[N, M]}\lambda_x\sigma_x^{(3)}.
\end{equation*}
Here,   $J_x$ is   the nearest neighbor coupling constant and $\lambda_x$ is  the strength of
an external magnetic field in direction $(3)$ at the site $x$. Throughout the paper we shall assume 
that $\{J_x\}_{x\in \zz}$ and $\{\lambda_x\}_{x\in \zz}$ are bounded sequences of real numbers and 
that $J_x\not=0$ for all $x\in\zz$. 

Consider the XY chain confined to the interval $\Lambda=[-M, M]$. Its Hamiltonian can be written as 
$$
H_\Lambda=H_l  + H_r +V,
$$
where $H_l$ is the Hamiltonian of the XY chain confined to $[-M, 0]$, $H_r$ is the Hamiltonian of the 
XY chain confined to $[1, M]$, and 
\[
V=\frac12 J_{0}\left(\sigma_{0}^{(1)}\sigma_{1}^{(1)}+\sigma_{0}^{(2)}\sigma_{1}^{(2)}\right),
\]
is the coupling energy.
Until the end of this section $\Lambda=[-M, M]$ is fixed and we shall omit the respective subscript,
\ie we write ${\cal H}= {\cal H}_\Lambda$, ${\cal O}={\cal O}_{\Lambda}$, $H= H_\Lambda$, 
{\sl etc.}\footnote{One can consider a more general setup where $H_{l/r}$ is the Hamiltonian of the 
XY chain confined to $[-M, -N-1]/[N+1, M]$, $H_{c}$ is the Hamiltonian of the XY chain confined to 
$[-N, N]$, and $H= H_l + H_{c} + H_r+ V$, where 
\[
V=\frac12 J_{-N-1}\left(
\sigma_{-N-1}^{(1)}\sigma_{-N}^{(1)}+\sigma_{-N-1}^{(2)}\sigma_{-N}^{(2)}
\right)+
\frac12 J_{N}\left(\sigma_{N}^{(1)}\sigma_{N+1}^{(1)}+\sigma_{N}^{(2)}\sigma_{N+1}^{(2)}\right).
\] 
In this scenario the left and the right parts of the chain are connected via the central part which is 
also an XY chain. In the thermodynamic limit one takes $M\rightarrow \infty$ while keeping $N$ fixed. 
The central part remains finite and represents a small system (quantum dot) connecting two  
infinitely extended chains (thermal reservoirs). All our results and proofs extend to this setting
\cite{La}. For notational simplicity, we will discuss here only the simplest case of two directly
coupled chains.}

The states of XY chain are described by density matrices on $\cH$. The expectation value of the
observable $A$ w.r.t.\;the state $\rho$ is $\rho(A)=\tr(\rho A)$ . We set 
\[ \tau^t(A)=\e^{\i t H}A\e^{-\i t H}.
\] In the Heisenberg/Schr\"odinger picture  the observables/states  evolve in time as 
\[A_t =\tau^t(A), \qquad \rho_t =\tau^{-t}(\rho).
\]
Obviously, $\rho_t(A)=\rho(A_t)$. The relative entropy  of the  state $\rho$ w.r.t.\;the state $\nu$ is 
defined by 
\[ 
S(\rho|\nu)= \tr (\rho(\log \nu - \log \rho)).
\]
We recall that $S(\rho|\nu)\leq 0$ and that $S(\rho|\nu)=0$ iff $\rho=\nu$.\footnote{
$S(\rho|\nu)=-\infty$ unless $\Ker \nu \subset \Ker \rho$.} For additional information about the
relative entropy we refer the reader to \cite{JOPP,OP}. 

Following the setup of \cite{AH}, we shall consider the evolution of the XY chain with an initial state 
given by
\begin{equation}
\omega=\frac{\e^{-\beta_l H_{l}-\beta_r H_{r}}}
{\tr(\e^{-\beta_l H_{l}-\beta_rH_{r}})},
\label{xy-initial}
\end{equation}
where $\beta_{l/r}>0$. 
Hence, the left part of the chain is initially in thermal equilibrium at inverse temperature $\beta_l$
while the right part of the chain is in thermal equilibrium at inverse temperature $\beta_r$. 
The XY chain is time-reversal invariant: there exists an anti-unitary involution
$\theta:\cH\rightarrow\cH$, described in Section~\ref{sec-jw},
such that, setting $\Theta(A)=\theta A\theta^{-1}$, one has
\[ 
\Theta(H_{l/r})=H_{l/r}, \qquad \Theta(V)= V,\qquad\Theta(\omega)=\omega.
\]

The observables describing the heat fluxes out of the left/right chain are 
\[
\Phi_{l/r }=\left. -\frac{\d\ }{\d t}\tau^t(H_{l/r })\right|_{t=0}=-\i [H, H_{l/r}]=\i [H_{l/r},V],
\]
and an easy computation gives 
\begin{align*}
\Phi_l &= \frac{1}{2} J_0 J_{-1} \sigma_0 ^{(3)} ( \sigma_1 ^{(1)} \sigma _{-1} ^{(2)} -\sigma_1 ^{(2)}
\sigma _{-1} ^{(1)} )
+ \frac{1}{2} J_0 \lambda _0 ( \sigma_1 ^{(2)} \sigma_0 ^{(1)} - \sigma_1 ^{(1)} \sigma_0
^{(2)}),\\[3mm]
\Phi_r&= \frac{1}{2} J_0 J_1 \sigma_1 ^{(3)} ( \sigma_0 ^{(1)} \sigma_2 ^{(2)} - \sigma_0 ^{(2)}
\sigma_2 ^{(1)} )
+ \frac{1}{2} J_0 \lambda _1 (\sigma_0 ^{(2)} \sigma_1 ^{(1)} - \sigma_0 ^{(1)} \sigma_1 ^{(2)}).
\end{align*}
The entropy production observable  of the XY chain is
\[ 
\sigma= -\beta_l \Phi_l - \beta_r \Phi_r.
\]
Note that $\Phi_{l/r}$ and $\sigma$  change sign under the time reversal:
\[ 
\Theta(\Phi_{l/r})=-\Phi_{l/r}, \qquad\Theta(\sigma)=-\sigma.
\]

The observable describing the mean entropy production rate over the time interval $[0, t]$ is 
\[
\Sigma^t= \frac{1}{t}\int_0^t \sigma_s \d s. 
\]
The basic properties of this observable are summarized in: 
\bep\label{prop-in-1}
\ben
\item $\log \omega_t = \log \omega + t\tau^{-t}(\Sigma^{t})$.
\item $S(\omega_t|\omega)= -t\omega(\Sigma^t)$.
\item $\tau^t(\Sigma^{-t})=\Sigma^t$.
\item $\Sigma^t =-\tau^t(\Theta(\Sigma^t))$.
In particular, $\sp(\Sigma^t)$ is symmetric w.r.t.\;zero and
$\dim\one_\phi(\Sigma^t)=\dim\one_{-\phi}(\Sigma^t)$ for all $\phi\in\sp(\Sigma^t)$. 
\een
\eep
The proof of Proposition \ref{prop-in-1} is elementary and can be found in \cite{JOPP}. 
Part (1) allows for the interpretation of the entropy production observable $\sigma$ as the 
{\em quantum phase space contraction rate}. Part (2) implies that for all $t>0$ the average 
entropy production rate over the interval $[0, t]$ is non-negative, \ie
\begin{equation}
\omega(\Sigma^t)=\sum_{\phi\in\sp(\Sigma^t)}\phi\,p^t_\phi\geq 0,
\label{sec-law-1}
\end{equation}
where 
\[
p^t_{\phi}=\omega(\one_\phi(\Sigma^t)),
\]
is the probability  that a measurement of $\Sigma^t$ in the state $\omega$ will yield the value $\phi$.
In particular, on average heat flows from the hotter to the colder part of the chain, in accordance 
with the (finite time) second law of thermodynamics. By Property (4), (\ref{sec-law-1}) is equivalent to
\[
\sum_{\substack{\phi\in\sp(\Sigma^t)\\\phi>0}}\phi(p_\phi^t - p^t_{-\phi})\geq 0.
\]
The finite time fluctuation relation is deeper. The direct quantization of the Evans-Searles
fluctuation relation in classical non-equilibrium statistical mechanics is the identity 
\begin{equation}
p_{-\phi}^t =\e^{-t\phi} p_\phi^t,
\label{es-naive-1}
\end{equation}
which should hold for all $\phi$ and $t>0$. An equivalent formulation of \eqref{es-naive-1} is that the 
functional 
\[ 
\ES_t(\alpha)= \log\omega\left(\e^{-\alpha t \Sigma^t}\right)
=\log\sum_{\phi\in\sp(\Sigma^t)} \e^{-\alpha t\phi}p_\phi^t,
\]
has the symmetry 
\begin{equation}
\ES_t(\alpha)=\ES_t(1-\alpha),
\label{es-naive-2}
\end{equation}
that holds for all  $\alpha \in \rr$ and $t>0$. The functional $\ES_t(\alpha)$ is the direct 
quantization of the finite time Evans-Searles functional in non-equilibrium classical statistical
mechanics \cite{JPR}. 
It is however easy to  show that  $\ES_t(1)>0=\ES_t(0)$ except at possibly countably many $t$'s  
and so the relations \eqref{es-naive-1} and \eqref{es-naive-2} {\em cannot} 
hold for all $t>0$ (see Exercise 3.3. in \cite{JOPP}). This point is further discussed in \cite{La}.

The first class of quantum entropic functionals that satisfy the Evans-Searles fluctuation relation for 
all times was proposed independently by Kurchan \cite{Ku} and Tasaki-Matsui \cite{TM}. They 
involve the fundamental concept of Full Counting Statistics associated to the repeated measurement 
protocol of the energy/entropy flow \cite{LL}. Consider the observable 
\[ 
{\cal S}=-\log \omega=\beta_l H_l+ \beta_rH_r + Z,
\]
where $Z= \log \tr (\e^{-\beta_lH_l-\beta_r H_r})$. Clearly, ${\cal S}_t=\tau^t(\cS)=-\log \omega_{-t}$, 
and Properties (1) and (3) imply
\[
\Sigma^t=\frac{1}{t}({\cal S}_t - {\cal S}).
\]
The probability that a measurement of ${\cal S}$ at time $t=0$ (when the system is in the state
$\omega$) yields $s\in \sp({\cal S})$ is
$ \omega(\one_s({\cal S}))$. After the measurement, the system is in the reduced state 
\[ 
\frac{\omega \one_{s}({\cal S})}{\omega (\one_{s}({\cal S}))},
\]
which evolves in time as 
\[ 
\e^{-\i t H} \frac{\omega \one_{s}({\cal S})}{\omega (\one_{s}({\cal S}))}\e^{\i t H}.
\]
A second  measurement of ${\cal S}$ at a later time $t>0$ yields $s^\prime\in\sp(\cS)$ with probability 
\[
\tr \left( \e^{-\i t H} \frac{\omega \one_{s}({\cal S})}{\omega (\one_{s}({\cal S}))}
\e^{\i tH}\one_{s^\prime}({\cal S})\right),
\]
and the joint probability of these  two measurements is 
\[
\tr \left( \e^{-\i t H} \omega \one_{s}({\cal S})\e^{\i t H}\one_{s^\prime}({\cal S})\right).
\]
The mean rate of entropy change between the two measurements is $\phi = (s^\prime -s)/t$
and its probability distribution is
\[ 
{\mathbb P}_t(\phi)= \sum_{s^\prime-s=t\phi}\tr \left( \e^{-\i t H} \omega \one_{s}({\cal S})
\e^{\i t H}\one_{s^\prime}({\cal S})\right).
\]
The discrete probability measure ${\mathbb P}_t$ is the {\em Full Counting Statistics} (FCS) for the
operationally defined entropy change over the time interval $[0, t]$ as specified by the above 
measurement protocol. Let 
\[ 
\FCS_t(\alpha)=\log \sum_{\phi}\e^{-\alpha t\phi}{\mathbb P}_t(\phi). 
\]
One easily verifies the identity 
\begin{equation}
\FCS_t(\alpha)=
\log \tr (\omega_t^{1-\alpha}\omega^\alpha),
\end{equation}
and time-reversal invariance implies  the fluctuation relation 
\begin{equation}
\FCS_t(\alpha)=\FCS_t(1-\alpha),
\label{sunday-sick}
\end{equation}
which holds for all $\alpha$ and $t$. The last relation  is equivalent to 
\begin{equation}
{\mathbb P}_t(-\phi)=\e^{-t \phi}{\mathbb P}_t(\phi).
\label{sunday-sick-1}
\end{equation}
The identities \eqref{sunday-sick} and \eqref{sunday-sick-1} provide a physically and mathematically
appealing extension of the Evans-Searles fluctuation relation to the quantum domain. 

Generalizations of the fluctuation relation \eqref{sunday-sick} have been recently proposed in
\cite{JOPP}. Note that
\[
\FCS_t(\alpha)=\log \tr \left(\e^{\frac{1-\alpha}{2}{\cal S}}
\e^{\alpha {\cal S}_t}\e^{\frac{1-\alpha}{2}{\cal S}}\right).
\]
For $p>0$ and $\alpha\in\rr$ we define the functionals $e_{p,t}(\alpha)$ by 
\begin{equation*}
e_{p,t}(\alpha)=
\begin{cases}
\log \tr \left(\left[ \e^{\frac{1-\alpha}{p}{\cal S}}\e^{\frac{2\alpha}{p} {\cal
S}_t}\e^{\frac{1-\alpha}{p}{\cal S}}\right]^{\frac{p}{2}}\right)&\text{if $0< p<\infty$},\\[3mm]
\log\tr \left(\e^{(1-\alpha){\cal S} +\alpha S_t}\right)&\text{if $p=\infty$}.
\end{cases}
\end{equation*}
Their basic properties are summarized in: 
\bep\label{pain}
\ben 
\item $e_{p, t}(0)=e_{p,t}(1)=0$.
\item The functions $\rr\ni \alpha \mapsto  e_{p ,t}(\alpha)$ are real-analytic and  convex.
\item $e_{p,t}(\alpha)=e_{p,-t}(\alpha)$.
\item $e_{p,t}(\alpha)=e_{p, t}(1-\alpha)$.
\item The functions $]0, \infty]\ni p \mapsto e_{p,t}(\alpha)$ are  continuous and  decreasing. 
\item $e_{p,t}^\prime(0)=\ES_t^\prime(0)=-e_{p, t}^\prime(1)=-t\omega(\Sigma^t)$. 
In particular, these derivatives do not depend on $p$.
\item $e_{2,t}(\alpha)=\FCS_t(\alpha)$ and 
\[
e_{2,t}^{\prime\prime}(0)=\ES_t^{\prime\prime}(0)=\int_0^t\int_0^t 
\omega\left((\sigma_s-\omega(\sigma_s))(\sigma_u -\omega(\sigma_u))\right)\d s\d u.
\]
\item $e_{\infty, t}(\alpha) = \max \left( S(\rho|\omega)- \alpha t \rho(\Sigma^t)\right)$,
where the maximum is taken over all states $\rho$. 
\een
\eep
{\bf Remark.} The variational characterization of $e_{\infty, t}(\alpha)$ in (8) is the quantization of
the variational characterization of the finite time Evans-Searles functional in classical non-equilibrium 
statistical mechanics. Regarding the physical interpretation
of $e_{p,t}(\alpha)$ for $p<\infty$, note that
\[
\tr\left(\e^{\frac{1-\alpha}{p}{\cal S}}\e^{\frac{2\alpha}{p} {\cal S}_t}\e^{\frac{1-\alpha}{p}
{\cal S}}\right)\big/\tr (\omega^{2/p})=
\int_\rr\e^{-\alpha t \phi} \d {\mathbb P}_{p,t}(\phi),
\]
where ${\mathbb P}_{p,t}$ is the FCS of the XY chain with scaled temperatures $2\beta_{l/r}/p$.
Restoring the scaling as 
\[
\tr \left(\left[\e^{\frac{1-\alpha}{p}{\cal S}}\e^{\frac{2\alpha}{p} {\cal S}_t}
\e^{\frac{1-\alpha}{p}{\cal S}}\right]^{p/2}\right),
\]
distorts the connection with the FCS but links $e_{p,t}(\alpha)$ with quantized Ruelle transfer 
operators which provide yet another way of quantizing  the classical Evans-Searles functional 
(see \cite{JOPP, JOP,JP3}).

For the proof of Proposition \ref{pain} we refer the reader to  \cite{JOPP}. The definitions and 
structural relations described in this section have a simple general algebraic origin and are
easily extended to any time-reversal invariant finite dimensional quantum system \cite{JOPP}. 
A similar remark applies to the thermodynamic limit results described in the next section
which hold for a much wider class of models than  XY chains and in particular for quite general 
lattice quantum spin systems (for example, for the open spin systems discussed in \cite{Ru2}). 
However, our study of the large time limit
$t\rightarrow \infty$ critically depends on the  specific properties of the XY model. 

\subsection{Thermodynamic limit}
\label{sec-thermlimit}
In this section we describe the  thermodynamic limit $\Lambda =[-M, M]\nearrow\zz$ (for  
discussion of the thermodynamic limit of  general spin systems we refer the reader to 
\cite{BR1,BR2,I,Ru1,S}). We use the subscript $M$ to denote the dependence of various
objects on the size of $\Lambda$ and write, for example, $\cH_M$, ${\cal O}_M$,
$H_M$, $e_{p,t,M}$, {\sl etc.} The $C^\ast$-algebra ${\cal O}$ of observables of the extended 
XY chain is the uniform closure of the algebra of local observables  
\[
{\cal O}_{\rm loc}=\bigcup_M {\cal O}_M,
\]
where we identify ${\cal O}_{M_1}$ with a subalgebra of ${\cal O}_{M_2}$ for $M_1<M_2$. 
For any $A\in {\cal O}_{\rm loc}$ the limit 
\[
\tau^t(A)=\lim_{M\rightarrow \infty}\e^{\i t H_M}A\e^{-\i t H_M},
\]
exists in norm and $\tau^t$ uniquely extends to a strongly continuous group of $\ast$-automorphisms
of ${\cal O}$ that describes the dynamics of the extended XY chain. The physical states of the 
extended XY chain are described by positive normalized linear functionals on ${\cal O}$. The 
expectation of an observable $A$ in the state $\rho$ is $\rho(A)$. In the Heisenberg/Schr\"odinger 
picture the observables/states evolve in time as 
\[
A_t= \tau^t(A), \qquad \rho_t =\rho \circ \tau^t.
\]
$S(\rho|\nu)$ denotes Araki's relative entropy of the state $\rho$ w.r.t.\;the state $\nu$ \cite{Ar1,Ar2} 
with the notational convention of \cite{BR2}. Let $\omega_{M}$ be the reference state 
\eqref{xy-initial} on ${\cal O}_M$. Then for all $A\in {\cal O}_{\rm loc}$ the limit
\[
\omega(A)=\lim_{M\rightarrow \infty}\omega_{M}(A),
\]
exists and $\omega$ uniquely extends to a state on ${\cal O}$ that describes the initial state of the
extended XY chain. The $C^\ast$ dynamical system $({\cal O}, \tau^t, \omega)$ 
describes the extended XY chain. This quantum dynamical system is time-reversal invariant, \ie 
there exists an anti-linear involutive $\ast$-automorphism $\Theta:{\cal O}\rightarrow {\cal O}$ such
that $\Theta\circ\tau^t=\tau^{-t}\circ\Theta$ for all $t$ and $\omega(\Theta(A))=\omega(A^\ast)$ 
for all $A\in {\cal O}$.

The observables $\Phi_{l/r}$ and $\sigma$  are obviously in ${\cal O}_{\rm loc}$. 
Let 
\begin{align*}
\Sigma^t&=\frac{1}{t}\int_0^t \sigma_s \d s,\\
\ES_t(\alpha)&=\log\omega(\e^{-\alpha t\Sigma^t}).
\end{align*}
We then have
\bep\label{therm1}
\ben 
\item 
\[
\lim_{M\rightarrow \infty} \Sigma^t_M=\Sigma^t,
\]
in norm and the convergence is uniform for $t$'s in compact sets. 
\item $S(\omega_t|\omega)=-t\omega(\Sigma^t).$
\item $\tau^t(\Sigma^{-t})=\Sigma^t$.
\item $\Sigma^t=-\tau^t(\Theta(\Sigma^t))$. In particular, $\sp(\Sigma^t)$ is symmetric with respect to
the origin.
\item 
\[
\lim_{M \rightarrow \infty}\ES_{t, M}(\alpha)=\ES_t(\alpha),
\]
and the convergence is uniform for $t$'s and $\alpha$'s in compact  sets. 
\een
\eep

Regarding the entropic functionals, we have:
\bep\label{therm2}
\ben
\item For all $\alpha\in\rr$ and $p\in]0,\infty]$ the limits 
\[
e_{p,t}(\alpha)=\lim_{M\rightarrow\infty}e_{p,t,M}(\alpha),
\]
exist and are finite.
\item $e_{p,t}(0)=e_{p,t}(1)=0$.
\item The functions $\rr\ni\alpha\mapsto e_{p,t}(\alpha)$ are real-analytic and  convex.
\item $e_{p,t}(\alpha)=e_{p,-t}(\alpha)$.
\item $e_{p,t}(\alpha)=e_{p,t}(1-\alpha)$.
\item The functions $]0,\infty]\ni p\mapsto e_{p,t}(\alpha)$ are continuous and decreasing. 
\item $e_{p,t}^\prime(0)=\ES_t^\prime(0)=-e_{p,t}^\prime(1)=-t\omega(\Sigma^t)$.
\item
\[
e_{2,t}^{\prime\prime}(0)=\ES_t^{\prime\prime}(0)=\int_0^t\int_0^t 
\omega\left((\sigma_s-\omega(\sigma_s))(\sigma_u-\omega(\sigma_u))\right)\d s\d u.
\]
\item As $M\to\infty$, the sequence ${\mathbb P}_{t,M}$ converges weakly towards a Borel 
probability measure ${\mathbb P}_t$ on $\rr$ and
\[
e_{2,t}(\alpha)=\log \int_{\rr}\e^{-t\alpha\phi}\d{\mathbb P}_t(\phi).
\]
All the moments of ${\mathbb P}_{t,M}$ converge to corresponding moments of ${\mathbb P}_t$. 
The measure ${\mathbb P}_t$ is the full counting statistics of the extended {\rm XY} chain.
\item $e_{\infty,t}(\alpha) = \sup\left(S(\rho|\omega)-\alpha t\rho(\Sigma^t)\right)$,
where the supremum is taken over all states $\rho$.
\een
\eep
The results of this section are  easy to prove (see Section \ref{sec-therm-lim}
and \cite{La}). 
Also, using standard techniques \cite{BR1,BR2,I,Ru1,S} these results are easily extended to 
other models in quantum statistical mechanics (general lattice quantum spin system, 
interacting fermionic models, Pauli-Fierz models, {\sl etc.}).
We emphasize, however,  that non-trivial thermodynamical behavior only emerges in the large 
time limit $t\rightarrow\infty$ and that controlling the large time limit of entropic functionals in
physically interesting models is typically a very hard analytic problem (see \cite{Ro, JOPP} for 
existing results and discussion of this point). The advantage of models like the XY chain is that 
the large time limit is easy to control via trace class scattering theory and very detailed 
information about limiting entropic functionals is available. In the next section we review some 
basic results of spectral and scattering theory that  we will need in this paper.  

\subsection{Operator theory preliminaries}
\label{sec-operator}
Let  $A$ be  a self-adjoint operator on a Hilbert space $\mathfrak{H}$. 
We denote by $\sp_{\ac}(A)$ the  absolutely 
continuous spectrum of $A$ and by  $\mathfrak{H}_\ac(A)$ the corresponding  
spectral subspace.  The projection on this  subspace is denoted by  $\one_{\ac}(A)$.  For any 
$\psi_1,\psi_2\in\mathfrak{H}$ the  boundary values 
\begin{equation*}
\langle \psi_1,(A-E-\i 0)^{-1}\psi_2\rangle=\lim_{\epsilon \downarrow 0} 
\langle \psi_1, (A-E-\i \epsilon)^{-1}\psi_2\rangle,
\end{equation*}
exist and are finite for Lebesgue a.e.\;$E\in \rr$. In what follows, whenever we write $\langle \psi_1,
(A-E-\i 0)^{-1}\psi_2\rangle$,
we will always assume that the limit exists and is finite. 
If $\psi_1=\psi_2=\psi$ then $\Im\langle\psi,(A-E-\i 0)^{-1}\psi \rangle\ge0$. If 
$\nu_\psi$ is the spectral measure of $A$ for $\psi$, then its absolutely continuous component with 
respect to Lebesgue measure is
\begin{equation}
\d\nu_{\psi,\ac}(E)=\frac{1}{\pi}\Im\langle\psi,(A-E-\i 0)^{-1}\psi \rangle\d E.
\label{monday-sick}
\end{equation}
For the proofs and references regarding the above results we  refer the reader to \cite{J}. 

The basic tool in virtually any study of XY chains is the Jordan-Wigner transformation \cite{JW} (see
also \cite{LSM, Ar3, AP, JOPP}).
This transformation associates to our XY chain  the Jacobi matrix
\[ 
(hu)(x) = J_x u(x+1) + \lambda_x u(x) + J_{x-1}u(x-1),
\]
(see Section \ref{sec-jw}).
As an operator on the Hilbert space $\ch=\ell^2(\zz)$, $h$ is bounded and self-adjoint. Set
\[
\fh_l= \ell^2(]-\infty, 0]), \qquad  \ch_r=\ell^2([1, \infty[),
\] 
and denote $h_{l/r}$ the restrictions of $h$ to $\ch_{l/r}$ with a Dirichlet boundary condition.
Note that $h= h_0 + v$ where  $h_0=h_l\oplus h_r$ and
\begin{equation}
v =J_0\left( |\delta_l\rangle \langle \delta_r| +|\delta_r\rangle \langle \delta_l|\right),
\label{def-v}
\end{equation}
$\delta_{l/r}$ being the Kronecker delta function at site $x=0/1$. We denote by $\nu_{l/r}$ the spectral 
measure of $h_{l/r}$ for $\delta_{l/r}$. By (\ref{monday-sick}), 
\[
\d\nu_{l/r,\ac}(E)=\frac{1}{\pi}F_{l/r}(E)\d E,
\]
where 
\[
F_{l/r}(E)=\Im G_{l/r}(E),\qquad
G_{l/r}(E)=\langle\delta_{l/r},(h_{l/r}-E-\i 0)^{-1}\delta_{l/r} \rangle.
\]
By the spectral theorem, one can identify $\fh_{\ac}(h_0)$ with $L^2(\rr, \d\nu_{l,\ac})\oplus
L^2(\rr, \d\nu_{r,\ac})$ and $h_0\upharpoonright\fh_{\ac}(h_0)$ with the operator of
multiplication by the variable $E\in \rr$. The set $\Sigma_{l/r,\ac}=\{ E\in \rr\,|\,F_{l/r}(E)>0\}$ is 
the essential support of the absolutely continuous spectrum of $h_{l/r}$. We set 
\[
{\cal E}=\Sigma_{l,\ac}\cap\Sigma_{r,\ac}.
\]

Finally, we recall a few basic facts that follow from trace class scattering theory \cite{RS,Y}. 
The wave operators
\[
w_{\pm}=\slim_{t\rightarrow \pm \infty}\e^{\i t h}\e^{-\i t h_0}\one_\ac(h_0),
\]
exist. The scattering matrix
\beq
s=w_+^\ast w_-
\label{smatrix}
\eeq
is unitary on $\fh_\ac(h_0)$ and acts as an operator
of multiplication by a unitary $2\times 2$ matrix function (called the on-shell scattering matrix) 
\[
s(E)=\left[\begin{array}{cc} s_{ll}(E) & s_{lr}(E) \\ s_{rl}(E) & s_{rr}(E)\end{array}\right],
\]
where, with $\bar l/\bar r=r/l$,
$$
s_{ab}(E)=\delta_{ab}+2\i J_0\left(
J_0\langle \delta_{\bar a}, (h-E-\i 0)^{-1}\delta_{\bar b}\rangle+\delta_{ab}-1
\right) \sqrt{F_a(E)F_b(E)}
$$
In our current setting these results can be easily proven  directly (see \cite{JKP, La}). 
Note that $s(E)$ is diagonal for Lebesgue a.e.\;$E\in \rr\setminus {\cal E}$. It follows from the formula
\[
\langle\delta_l,(h-E-\i0)^{-1}\delta_r\rangle
=\langle\delta_r,(h-E-\i0)^{-1}\delta_l\rangle
=-\frac{J_0G_l(E)G_r(E)}{1-J_0^2G_l(E)G_r(E)}
\]
that  $s(E)$ is symmetric and not diagonal for Lebesgue 
a.e.\;$E\in{\cal E}$.

The Jacobi matrix $h$ is called {\em reflectionless} iff $s(E)$ is off-diagonal for $E\in {\cal E}$.
In other words, $h$ is reflectionless if the transmission probability satisfies $|s_{lr}(E)|^2=1$ for 
$E\in {\cal E}$. For other equivalent
definitions of reflectionless we refer the reader to Chapter 8 in \cite{Te} (see also \cite{La} for a
discussion). For additional information and references about reflectionless Jacobi matrices we refer the
reader to the recent work \cite{Re}.

\subsection{The large time limit: NESS}
\label{sec-ness}

The basic result concerning the existence of a non-equilibrium steady state (NESS) of the
XY chain is:

\bet\label{ness1}
Suppose that $h$ has purely absolutely continuous spectrum. Then for all $A\in {\cal O}$ the 
limit
\[
\langle A\rangle_+=\lim_{t\rightarrow \infty} \omega(\tau^t(A)),
\]
exists. The state $\omega_+(\,\cdot\,)=\langle\,\cdot\,\rangle_+$ is called the NESS of the quantum
dynamical system $({\cal O}, \tau^t, \omega)$. The steady  state heat fluxes are 
\begin{equation}
\langle \Phi_{l} \rangle_+=-\langle \Phi_{r} \rangle_+
=\frac{1}{4\pi}\int_{\cal E}
E|s_{lr}(E)|^2\frac{\sinh\left(\Delta\beta E/2\right)}{\cosh \left( 
\beta_rE/2\right)\cosh \left(\beta_lE/2\right)}\d E,
\label{LB-form-2}
\end{equation}
and the steady state entropy production is 
$$
\langle \sigma \rangle_+=-\beta_l\langle \Phi_l\rangle_+ -\beta_r \langle \Phi_r\rangle_+
=\Delta\beta\langle \Phi_{l} \rangle_+,
$$
where $\Delta\beta=\beta_r-\beta_l$.
\eet

With only notational changes the proof in \cite{AP} extend to the proof of Theorem
\ref{ness1}\footnote{Alternatively, Theorem \ref{ness1} can established by applying the results of 
\cite{JKP} to the Jordan-Wigner transformed XY chain.}. If in addition to its  absolutely 
continuous spectrum $h$ has non-empty  pure point spectrum then one can show that the limit 
\[
\langle A\rangle_+= \lim_{T\rightarrow \infty}\frac{1}{T}\int_0^T\omega(\tau^s(A))\d s,
\] 
exists for all $A\in {\cal O}$ and that the formula \eqref{LB-form-2} remains valid (see \cite{AJPP2}). 
In presence of singular continuous spectrum the existence of a NESS is not known. 

If $|{\cal E}|$, the Lebesgue measure of ${\cal E}$, is zero, then obviously there is no energy transfer
between the left and the right part of the chain and 
$\langle\sigma\rangle_+=\langle \Phi_{l/r}\rangle_+=0$ for all $\beta_{l/r}$. If $|{\cal E}|>0$, then
$\langle \sigma\rangle_+ >0$ iff $\beta_l\not=\beta_r$. Moreover, $\langle \Phi_l\rangle_+>0$ if
$\beta_l<\beta_r$ and $\langle \Phi_l\rangle_+<0$ if $\beta_l>\beta_r$. The formula \eqref{LB-form-2}
is of course a special case of the celebrated Landauer-B\"uttiker formula that expresses currents in 
terms of the scattering data.\footnote{ We refer the reader to \cite{AJPP2,N,JKP} for mathematically 
rigorous results regarding the Landauer-B\"uttiker formula and for references to the vast physical 
literature on the subject.} 

Fluctuations of the mean entropy production rate $\Sigma^t$ with respect to the NESS $\omega_+$ 
are controlled by the functional 
\[
\GC_t(\alpha)=\log \omega_+(\e^{-\alpha t \Sigma^t}),
\]
which is the direct quantization of the Gallavotti-Cohen functional in non-equilibrium
classical statistical mechanics \cite{JPR}.

\subsection{The large time limit: Entropic functionals}
\label{sec-timelimit}

In this section we state our main results. It is convenient to introduce the following matrix notation.
We let
\[
k_0(E)=\left[\begin{array}{cc} -\beta_lE & 0 \\ 0 & -\beta_rE\end{array}\right],
\]
set
\beq
K_\alpha(E)=\e^{k_0(E)/2} \e^{\alpha(s^\ast(E)k_0(E)s(E)-k_0(E))}\e^{k_0(E)/2},
\label{Kalpha}
\eeq
and note that $K_0(E)=\e^{k_0(E)}$. We further set
\beq
K_{\alpha,p}(E)
=\left[\e^{k_0(E)(1-\alpha)/p}s(E)\e^{k_0(E)2\alpha/p}s^\ast(E)\e^{k_0(E)(1-\alpha)/p}\right]^{p/2},
\label{Kalphap}
\eeq
for $p \in ]0, \infty[$, and
\beq
K_{\alpha,\infty}(E)=\lim_{p\rightarrow \infty} K_{\alpha,p}(E)
=\e^{(1-\alpha)k_0(E) + \alpha s(E)k_0(E)s^\ast(E)}.
\label{Kalphainfty}
\eeq
Note that $K_{0,p}(E)= K_0(E)$. Moreover, it is a simple matter to check that, for $E,\alpha\in\rr$
and $p\in]0,\infty]$,
$$
K_{\alpha}(E)=K_0(E)\Leftrightarrow
K_{\alpha,p}(E)=K_0(E)\Leftrightarrow
[s(E),k_0(E)]=0\Leftrightarrow
\beta_l=\beta_r \text{ or }s(E) \text{ is diagonal.}
$$
Recall that the last condition holds for Lebesgue a.e.\;$E\in\rr\setminus{\cal E}$, but fails for
Lebesgue a.e.\;$E\in{\cal E}$.

\bet\label{mainthm}
Suppose that $h$ has purely absolutely continuous spectrum. Then the following holds:
\ben 
\item For $\alpha\in\rr$ and $p\in ]0, \infty]$, 
\[
e_{p,+}(\alpha)=\lim_{t\rightarrow\infty}\frac{1}{t}e_{p,t}(\alpha)
=\int_{\cal E}\log\left(\frac{\det(1+ K_{\alpha,p}(E))}{\det(1+K_{0,p}(E))}\right)\frac{\d E}{2\pi}, 
\]
\[
\lim_{t\rightarrow\infty}\frac{1}{t}\ES_t(\alpha)
=\lim_{t\rightarrow\infty}\frac{1}{t}\GC_t(\alpha)
=e_{+}(\alpha)
=\int_{\cal E}\log\left(\frac{\det (1+K_{\alpha}(E))}{\det (1+ K_0(E))}\right)\frac{\d E}{2\pi}.
\]
These functionals are identical to zero iff $|{\cal E}|=0$ or $\beta_l=\beta_r$. In what follows we
assume that $|{\cal E}|>0$ and $\beta_l\not=\beta_r$. 
\item The functions $\rr\ni\alpha\mapsto e_{p,+}(\alpha)$ are real-analytic and strictly convex.
Moreover, $e_{p,+}(0)=0$, $e_{p,+}^\prime(0)=-\langle\sigma\rangle_+$, and
\begin{equation*}
e_{p,+}(\alpha)=e_{p,+}(1-\alpha).
\end{equation*}
\item The function $\rr\ni\alpha\mapsto e_+(\alpha)$ is real-analytic and strictly convex.
Moreover, it satisfies $e_+(0)=0$, $e_{+}^\prime(0)=-\langle \sigma \rangle_+$, and 
\begin{equation}
e_{+}^{\prime\prime}(0)=e_{2,+}^{\prime\prime}(0)
=\lim_{T\rightarrow\infty}\frac{1}{T}\int_0^T\left\{
\frac12\int_{-t}^t\left<(\sigma_s-\langle\sigma\rangle_+)(\sigma-\langle\sigma\rangle_+)
\right>_+\d s\right\}\d t.
\label{sat-ti}
\end{equation}
\item 
$e_+(1)>0$ unless $h$ is reflectionless.  If $h$ is reflectionless then 
\[
e_{+}(\alpha)=\int_{\cal E}\log\left(\frac{\cosh((\beta_l(1-\alpha)+\beta_r\alpha)E/2)
\cosh((\beta_r(1-\alpha)+\beta_l\alpha)E/2)}{\cosh(\beta_l E/2)\cosh (\beta_r E/2)}\right)\frac{\d E}{2\pi},
\]
and  $e_+(\alpha)=e_+(1-\alpha)$.
\item The function $]0,\infty]\ni p\mapsto e_{p,+}(\alpha)$ is continuous and decreasing. It is
strictly decreasing for $\alpha\not\in\{0,1\}$ unless $h$ is reflectionless. If $h$ is reflectionless, 
then $e_{p,+}(\alpha)$ does not depend on $p$ and is equal to  $e_+(\alpha)$. 
\een
\eet

{\bf Remark 1.} Typically, the correlation function
$C(s)=\left<(\sigma_s-\langle\sigma\rangle_+)(\sigma-\langle\sigma\rangle_+)\right>_+$
is not absolutely integrable and the Ces\'aro sum in \eqref{sat-ti} cannot be replaced by the integral
of $C(s)$ over $\rr$ without further assumptions.

{\bf Remark 2.} As already noticed, time-reversal invariance implies that the on-shell scattering matrix 
is symmetric. It follows that one can replace $s(E)$ by $s^\ast(E)$ and $s^\ast(E)$ by $s(E)$
in \eqref{Kalpha}--\eqref{Kalphainfty} when inserted in the formulas of Part (1).

{\bf Remark 3.} The vanishing of entropic functionals at $\alpha=1$ is called Kawasaki's identity,
see \cite{CWWSE}. Thus, the Kawasaki identity holds for all $e_{p}$'s and fails for $e_+$ unless $h$
is reflectionless. 
\bigskip

Theorem \ref{mainthm} has the following implications. To avoid discussing trivialities until the end of
this section we assume that $|{\cal E}|>0$ and $\beta_l\not=\beta_r$.  Recall that 
\[
e_{2,t}(\alpha)=\FCS_t(\alpha)=\int_\rr\e^{-\alpha t \phi}\d {\mathbb P}_t(\phi),
\]
where ${\mathbb P}_t$ is the FCS measure of the  extended XY chain. If ${\mathbb P}_{{\rm ES},t}$
and ${\mathbb P}_{{\rm GC},t}$ are respectively the spectral measures of $\Sigma^t$ for $\omega$
and $\omega_+$ we also have 
\[
\ES_t(\alpha)=\log\int_\rr \e^{-\alpha t\phi}\d {\mathbb P}_{{\rm ES},t}(\phi), \qquad 
\GC_t(\alpha)=\log\int_\rr \e^{-\alpha t\phi}\d {\mathbb P}_{{\rm GC},t}(\phi).
\]
The respective large deviation rate functions are given by 
\begin{align*}
I_{{\rm FCS}+}(\theta)&=-\inf_{\alpha \in \rr}(\alpha \theta+e_{2,+}(\alpha)),\\[4pt]
I_+(\theta)&=-\inf_{\alpha \in \rr}(\alpha \theta+e_+(\alpha)).
\end{align*}
The functions $I_{+}(\theta)$ and $I_{{\rm FCS}+}(\theta)$ are non-negative, real-analytic, strictly 
convex and vanish at the single point $\theta=\langle\sigma\rangle_+$. These two rate functions are 
different unless $h$ is reflectionless. The symmetry $e_{2,+}(\alpha)=e_{2,+}(1-\alpha)$ implies 
\[ 
I_{{\rm FCS}+}(\theta)=  I_{{\rm FCS}+}(-\theta) +\theta.
\]
The rate function $I_+(\theta)$ satisfies this relation only if $h$ is reflectionless. Theorem
\ref{mainthm} implies:
\begin{corollary}\label{mainthm-1}
Suppose that $h$ has purely absolutely continuous spectrum. 
\ben 
\item The Large Deviation Principle holds:  for any open set $O\subset\rr$,
\[
\lim_{t\rightarrow\infty}\frac{1}{t}\log{\mathbb P}_{{\rm ES},t}(O)
=\lim_{t\rightarrow\infty}\frac{1}{t}\log{\mathbb P}_{{\rm GC},t}(O)
=-\inf_{\theta\in O}I_+(\theta),
\] 
\[
\lim_{t\rightarrow \infty}\frac{1}{t}\log {\mathbb P}_{{\rm FCS},t}(O)
=-\inf_{\theta \in O}I_{{\rm FCS}+}(\theta).
\]
\item The Central Limit Theorem holds: for any Borel set $B\subset\rr$, let
$B_t=\{\phi\,|\,\sqrt{t}(\phi-\langle\sigma\rangle_+)\in B\}$. Then
\[
\lim_{t\rightarrow\infty}{\mathbb P}_{{\rm ES},t}(B_t)
=\lim_{t\rightarrow\infty}{\mathbb P}_{{\rm GC},t}(B_t)
=\lim_{t\rightarrow\infty}{\mathbb P}_{{\rm FCS},t}(B_t)
=\frac{1}{\sqrt{2\pi D_+}}\int_B\e^{-\phi^2/2D_+}\d\phi,
\]
where the variance is $D_+=e_+^{\prime\prime}(0)$.
\een
\end{corollary}


\subsection{Remarks}
\label{sec-discussion}
Due to non-commutativity it is natural that non-equilibrium quantum statistical mechanics has a richer
mathematical structure than its classical counterpart. This is partly reflected in the emergence of 
novel entropic functionals. The direct quantizations of the Evans-Searles and Gallavotti-Cohen 
entropic functionals typically will not have the symmetries which, starting with the seminal works 
\cite{CG, ES}, have played a central role in recent developments in non-equilibrium classical statistical 
mechanics. The mathematical theory of entropic fluctuations in quantum statistical mechanics is
an emerging research direction and our testing of the existing structural results on the specific
example of the XY chain has lead to some surprising (at least to us) results. 

If  in a given model all the functionals $e_{p,+}(\alpha)$ are equal and coincide with $e_+(\alpha)$ 
we shall say that in this model the entropic identity holds. For the XY chain we have shown that the 
entropic identity holds iff the underlying Jacobi matrix is reflectionless. Note that the large time 
fluctuations of ${\mathbb P}_{{\rm ES},t}$ and ${\mathbb P}_{{\rm GC},t}$ are identical for all
XY chains (\ie for any $h$) and this is not surprising---the same phenomenon generally holds in 
classical/quantum non-equilibrium statistical mechanics (this is the {\em principle of regular entropic 
fluctuations} of \cite{JPR, JOP}). On the other hand, the full counting statistics of repeated 
measurement of entropy flux is obviously of quantum origin with no classical counterpart
and in general one certainly does not expect that the large time fluctuations of 
${\mathbb P}_{{\rm FCS},t}$ are equal to those of ${\mathbb P}_{{\rm ES}/{\rm GC},t}$. In the 
same vein, one certainly expects that $e_{\infty,+}(\alpha)<e_{2,+}(\alpha)$ for $\alpha \not=0,1$, 
and more generally, that all functionals $e_{p,+}(\alpha)$ are different.

Needless to say, there are very few models for which the existence of $e_{p,+}(\alpha)$ and
$e_{+}(\alpha)$ can be rigorously proven and the phenomenon of entropic identity, to the best of our
knowledge, has not been previously discussed in mathematically rigorous literature on the subject. The
XY chain is instructive since the functionals $e_{p,+}(\alpha)$ and $e_{+}(\alpha)$ can be computed in
closed form and the  entropic identity of the model can be identified with the reflectionless property of the underlying Jacobi matrix.
The question whether such a physically and mathematically natural result can be extended beyond 
exactly solvable models like the XY chain or the Electronic Black Box Model\footnote{See 
\cite{AJPP1, AJPP2, JKP, JOPP} for discussion of these models and references.}
remains to be studied in the future.

\subsection{Examples}
\label{sec-reflectionless}

{\em Schr\"odinger case.} If $J_x= J$ for all $x$ then the Jacobi matrix $h$ is the discrete
Schr\"odinger operator
\[
(hu)(x)=J (u(x+1) + u(x-1)) +\lambda_x u(x).
\]
In the literature, the  most commonly studied   XY chains  correspond to this case. The assumption
that $h$ has purely absolutely continuous spectrum plays a critical role in the formulation and the 
proof of Theorem \ref{mainthm}. The striking fact is that the only known examples of discrete 
$d=1$ Schr\"odinger operators with purely absolutely continuous spectrum are reflectionless. If  
the potential $\lambda_x$ is constant or periodic then $h$ has purely absolutely continuous spectrum
(and is reflectionless). The only other known examples involve quasi-periodic potentials. For example, 
if $\lambda_x=\cos(2\pi \alpha x+\theta)$, then for $|J|>1/2$ the operator $h$ has purely absolutely 
continuous spectrum and is reflectionless (see \cite{Av}). In this context it is an important open
problem whether there exists a one-dimensional discrete Schr\"odinger operator $h$ which is 
reflectionless and has purely absolutely continuous spectrum and therefore an XY chain with 
$J_x$ constant for which the entropic identity fails .\footnote{It  is believed that such an example exists. 
We are grateful to C.~Remling and B.~Simon for discussions on reflectionless Jacobi matrices.}

{\em Jacobi case.} If the $J_x$ are allowed to vary, then it is easy to produce examples of $h$ which
are not reflectionless and have purely absolutely continuous spectrum (take $\lambda_x=0$, $J_x=1$
for $x>0$ and $J_x=1/2$ for $x\leq 0$). In fact, the vast majority of Jacobi matrices with purely 
absolutely continuous spectrum are  not reflectionless. To illustrate this point, let 
$q: [-2,2 ]\rightarrow \cc $ be such that $|q(E)|\leq 1$ and 
\[
\int_{-2}^{2} \frac{\log |q(E)|}{\sqrt{2-E^2}}\d E <\infty.
\]
Then there exists a Jacobi matrix $h$ such that the spectrum of $h$ is purely absolutely continuous 
and equal to $[-2,2]$, and such that $s_{lr}(E)= q(E)$ for Lebesgue a.e.\;$E\in [-2,2]$  \cite{VY}.
The only reflectionless Jacobi matrix in this class is the Schr\"odinger operator with $J=\pm 1$ and 
$\lambda_x=0$ (see \cite{Te}). A similar result holds if the interval $[-2,2]$ is replaced by a 
homogeneous set (say, a finite union of intervals or even a Cantor set of positive measure) \cite{VY}. 

\section{Proofs}
\label{sec-proofs}

\subsection{The Jordan-Wigner transformation}
\label{sec-jw}

Consider the XY chain confined to $\Lambda=[-M, M]$. In this subsection $M$ is fixed and we 
drop the respective subscript. Let $\Lambda_l=[-M, 0]$, $\Lambda_r=[1, M]$. Set  
$\ch=\ell^2(\Lambda)$, $\ch_{l/r}=\ell^2(\Lambda_{l/r})$,  and let $h$, $h_{l/r}$  be the restrictions of 
the Jacobi matrix  to $\Lambda$, $\Lambda_{l/r}$. Whenever the meaning is clear we extend 
$h_{l/r}$ to $\ch$ by $h_l \oplus 0/0\oplus h_r$. Let $h_0=h_l+h_r$ and 
\[
h = h_0 +v,
\]
where $v$ is given by \eqref{def-v}.

We shall assume that the reader is familiar with the formalism of the fermionic second quantization 
(see \cite{BR2, BSZ, AJPP1} for general results and \cite{JOPP}  for a pedagogical introduction to 
this topic). ${\cal F}$ denotes the antisymmetric Fock space over $\ch$ and $a^\ast(f)/a(f)$ the 
creation/annihilation operator associated to $f\in\ch$. We write $a^\ast(\delta_x) =a^\ast_x$ {\sl etc.} 
Set $S_{-M}=1$ and
\[
S_x=\prod_{y\in [-M, x[}(2a_y^\ast a_y -1),
\]
for $x\in ]-M, M]$. 
 
\bep\label{prop-jw}
There exists a unitary $U_{\rm JW}:{\cal H}\rightarrow{\cal F}$, called the Jordan-Wigner
transformation, such that
\[
U_{\rm JW}\sigma_x^{(1)}U_{\rm JW}^{-1} = S_x(a_x + a_x^\ast), \qquad 
U_{\rm JW}\sigma_x^{(2)}U_{\rm JW}^{-1} = \i S_x(a_x - a_x^\ast), \qquad 
U_{\rm JW}\sigma_x^{(3)}U_{\rm JW}^{-1} = 2a_x^\ast a_x-1.
\]
\eep

The Jordan-Wigner transformation goes back to \cite{JW} and  the proof 
of the above proposition is well-known (a pedagogical exposition can be  found in \cite{JOPP}). 
An immediate consequence  are the identities
\[
U_{\rm JW}H_{l/r}U_{\rm JW}^{-1}=\d\Gamma(h_{l/r}), \qquad 
U_{\rm JW}VU_{\rm JW}^{-1}=\d\Gamma(v), \qquad
U_{\rm JW}H U_{\rm JW}^{-1}=\d\Gamma(h),
\]
\begin{align*}
U_{\rm JW}\Phi_{l}U_{\rm JW}^{-1}
&=-\i J_0J_{-1}\left(a^\ast_1 a_{-1}-a^\ast_{-1} a_1\right)
-\i J_0\lambda_0\left(a^\ast_1 a_0- a^\ast_0 a_1\right), \\[3mm]
U_{\rm JW}\Phi_{r}U_{\rm JW}^{-1}
&=-\i J_0J_{1}\left(a^\ast_0 a_2-a^\ast_2 a_0\right)
-\i J_0\lambda_1\left(a^\ast_0 a_1- a^\ast_1 a_0\right).
\end{align*}
It is also easy to see that the confined XY chain is time-reversal invariant with $\theta=
U_{\rm JW}^{-1}\Gamma( j)U_{\rm JW}$, where $j\psi=\bar\psi$ is the standard complex conjugation 
on $\ell^2(\Lambda)$.

In what follows we will work only in the fermionic representation of the confined XY chain. With a
slight abuse of notation we write $H_{l/r}=\d\Gamma(h_{l/r})$,  $V=\d\Gamma(v)$, 
$H = H_l + H_r + V=\d\Gamma(h)$, {\sl etc.} 

\subsection{Basic formulas}

In this section $\Lambda=[-M, M]$ is again fixed and we drop the respective subscript. We shall make
repeated use of the identity
\[
\tr (\Gamma(A))=\det (1+A),
\]
which holds for any linear map $A:\ch\rightarrow\ch$. Set 
\[
k=-\beta_l h_l -\beta_rh_r, \qquad k_t =\e^{\i t h}k\e^{-\i t h}.
\]
The initial state of the system is described by the density matrix 
\[
\omega =\frac{\e^{-\beta_l H_l-\beta_r H_r}}{\tr(\e^{-\beta_l H_l -\beta_r H_r})}
=\frac{\Gamma(\e^k)}{\det (1+\e^k)}.
\]
Since 
\[
\left(\omega^{(1-\alpha)/p}\omega_{t}^{2\alpha/p}\omega^{(1-\alpha)/p}\right)^{p/2}=
\frac{\Gamma\left(\left(\e^{(1-\alpha)k/p}\e^{2\alpha k_{-t}/p}\e^{(1-\alpha)k/p}\right)^{p/2}\right)}
{\det(1+\e^k)},
\]
we have 
\[
e_{p,t}(\alpha)=\log\frac{\det\left(1+\left(
\e^{(1-\alpha)k/p}\e^{2\alpha k_{-t}/p}\e^{(1-\alpha)k/p}\right)^{p/2}\right)}
{\det (1+\e^k)},
\]
for $0<p<\infty$. Similarly, 
\[
e_{\infty,t}(\alpha)=\log\frac{\det\left(1+\e^{(1-\alpha)k+\alpha k_{-t}}\right)}
{\det (1+\e^k)},
\]
and 
\[
\ES_t(\alpha)=\log\frac{\det\left(1+\e^{k/2}\e^{\alpha(k_t-k)}\e^{k/2}\right)}{\det (1+\e^k)}.
\]

Clearly, the functions $\alpha\mapsto e_{p,t}(\alpha)$, $\alpha\mapsto\ES_t(\alpha)$ are real
analytic. For $u\in\rr$ and $0<p<\infty$  set 
\begin{equation}
{\cal K}_{p,t}(\alpha, u)=\frac{1}{2} \e^{-(1-\alpha)k_{tu}/p}\left(1+
\left(\e^{(1-\alpha)k_{tu}/p}\e^{2\alpha k_{-t(1-u)}/p}
\e^{(1-\alpha)k_{tu}/p}\right)^{-p/2}\right)^{-1}\e^{(1-\alpha)k_{tu}/p}  + {\rm h.c.},
\label{kp}
\end{equation}
where  h.c.\;stands for the hermitian conjugate of the first term. We also set
 \begin{equation}
 {\cal K}_{\infty,t}(\alpha, u) = \left(1 + \e^{-(1-\alpha)k_{tu}-\alpha k_{-t(1-u)}}\right)^{-1},
 \label{kinf}
 \end{equation}
 and 
 \begin{equation}
{\cal K}_{{\rm ES},t}(\alpha, u)=-\left(1 + \e^{-\alpha (k_{t(1-u)} -
k_{-tu})}\e^{-k_{-tu}}\right)^{-1}.
 \label{kes}
 \end{equation}
 The following lemma will play the key role in the sequel.
\bel\label{lemma-key}
\begin{enumerate}[{\rm (1)}]
\item For $p\in]0,\infty]$,
\beq
e_{p,t}(\alpha)=t\int_0^\alpha\d\gamma\int_0^1\d u\,\tr\left({\cal K}_{p,t}(\gamma,u)\i[k,h]\right).
\label{der-form-1}
\eeq
\item
\beq
\ES_t(\alpha)=t\int_0^\alpha\d\gamma\int_0^1\d u\,\tr\left({\cal K}_{{\rm ES},t}(\gamma,u)\i[k,h]\right).
\label{der-form-2}
\eeq
\end{enumerate}
\eel
To prove the lemma we need the following preliminary result (for the proof see Lemma 2.2. in 
\cite{HP}):

\bel\label{hiai-petz} 
Let $F$ be a differentiable function of the real variable $\alpha$ taking values in Hermitian strictly
positive matrices on $\ch$. Then for any $p\in ]0,\infty[$, 
\[
\frac{\d\ }{\d\alpha}\tr \log (1+F(\alpha)^p)
=p\,\tr\left((1+ F(\alpha)^{-p})^{-1}F(\alpha)^{-1}\frac{\d F(\alpha)}{\d\alpha}\right).
\]
\eel
 
{\bf Proof of Lemma \ref{lemma-key}.}
We will derive \eqref{der-form-2}. Relation \eqref{der-form-1} can be obtained in a similar way. 
Note that
\[
\ES_t(\alpha)=\tr\log(1+F(\alpha))-\log\det(1+\e^k),
\]
where 
\[
F(\alpha)=\e^{k/2}\e^{\alpha(k_t-k)}\e^{k/2}.
\]
Lemma \ref{hiai-petz} and simple algebra yield 
\[
\frac{\d\ }{\d \alpha}\ES_t(\alpha)=\tr\left(\left(1+\e^{-\alpha(k_t-k)}\e^{-k}\right)^{-1}(k_t-k)\right).
\]
Since 
\[
k_t- k=\int_0^1\frac{\d\ }{\d u}k_{tu}\,\d u=t\int_0^1\e^{\i tuh}\i[h,k]\e^{-\i tuh}\d u,
\]
we have 
\[
\frac{\d\ }{\d \alpha}\ES_t(\alpha)=-t\int_0^1\tr\left(\e^{-\i tuh}\left(1+\e^{-\alpha(k_t-k)}
\e^{-k}\right)^{-1}\e^{\i uth}\i[k,h]\right)\d u.
\]
Using $\ES_t(0)=0$, integration yields Relation \eqref{der-form-2}. \qed

\subsection{Thermodynamic limit} 
\label{sec-therm-lim}

The proof of Proposition \ref{therm1} is standard and we shall omit it (see \cite{La}). To prove 
Proposition \ref{therm2}, we consider $h_{M}$ as an operator on $\ch=\ell^2(\zz)$ ($h_M$ acts
as zero on the orthogonal complement of $\ch_\Lambda$ in $\ch$). Clearly, $h_M\rightarrow h$
strongly and so the strong limits
\[
\slim_{M\rightarrow\infty} {\cal K}_{p,t,M}(\alpha,u)={\cal K}_{p,t}(\alpha,u), \qquad 
\slim_{M\rightarrow\infty} {\cal K}_{{\rm ES},t,M}(\alpha,u)={\cal K}_{{\rm ES},t}(\alpha,u),
\]
exist. Moreover, ${\cal K}_{p,t}(\alpha, u)$ and ${\cal K}_{{\rm ES},t}(\alpha,u)$ are given by the 
formulas \eqref{kp}, \eqref{kinf}, \eqref{kes} and hence are norm continuous functions of
$(p,t,\alpha,u)$. Since $\i[k_M,h_M]=\i[k,h]$ does not depend on $M$ and is a finite rank operator, 
an application of dominated convergence yields 
\beq
e_{\#,t}(\alpha)=\lim_{M\rightarrow\infty}e_{\#, t, M}(\alpha)
=t\int_0^\alpha\d\gamma\int_0^1\d u\,\tr\left({\cal K}_{\#,t}(\gamma,u)\i[k,h]\right),
\label{der-form-td}
\eeq
where $\#$ stands for $p$ or ES and $e_{{\rm ES}, t}(\alpha)={\rm ES}_t(\alpha)$. The functions $\e_{\#,t}(\alpha)$ are jointly continuous in 
$(p,t,\alpha)$ and real-analytic in $\alpha$. The Trotter product formula yields
\[
\lim_{p\rightarrow\infty}e_{p,t}(\alpha)=e_{\infty,t}(\alpha),
\]
and Parts (1)--(6) of Proposition \ref{therm2} follow. To prove (7) and (8), note that for given 
$\#$ in $\{p, {\rm ES}\}$ and $t$ there is an $\epsilon >0$ such that $e_{\#,t,M}(\alpha)$ extends 
analytically to the ball $|\alpha|<\epsilon$ in $\cc$ and satisfies 
$\sup_{M>0,|\alpha|<\epsilon}|e_{\#,t,M}(\alpha)|<\infty$. This observation, Parts (6) and (7) of
Proposition \ref{pain} and Vitali's convergence theorem (see, for example, Appendix B in \cite{JOPP}) 
imply (7) and (8). To prove (9), we note that the function
\begin{align*}
\alpha\mapsto\int\e^{-\alpha t\phi}\,\d{\mathbb P}_{t,M}(\phi)
&=\e^{e_{2,t,M}(\alpha)}
=\frac{\det\left(1+\e^{(1-\alpha)k_M}\e^{-\i th_M}\e^{\alpha k_M}\e^{\i th_M}\right)}
{\det (1+\e^{k_M})}\\[4pt]
&=\det\left(1+(1+\e^{-k_M})^{-1}
(\e^{-\alpha k_M}\e^{-\i th_M}\e^{\alpha k_M}\e^{\i th_M}-1)\right),
\end{align*}
is entire analytic. The bound $|\det(1+A)|\le\e^{\|A\|_1}$, where $\|A\|_1$ denotes the
trace norm of $A$, together with the formula
$$
\e^{-\alpha k_M}\e^{-\i th_M}\e^{\alpha k_M}\e^{\i th_M}-1=
\int_0^t\e^{-\alpha k_M}\e^{-\i sh_M}\i[\e^{\alpha k_M},v]\e^{\i sh_M}\,\d s,
$$
(where $v$ is finite rank) imply that for any bounded set $B\subset\cc$, 
\[
\sup_{\alpha\in B,M>0}\left|\e^{e_{2,t,M}(\alpha)}\right| <\infty.
\]
By Vitali's convergence theorem the sequence of characteristic functions of the measures 
${\mathbb P}_{t,M}$ converges locally uniformly towards an entire analytic function.
The existence of the weak limit ${\mathbb P}_t$ follows (see e.g., Corollary 1 to Theorem 26.3 
in \cite{Bi}) and the convergence of the moments is a direct consequence of Vitali's theorem.
Finally, (10) is a general fact which follows from Araki's perturbation 
theory of modular structure, see \cite{JOP} for the proof and additional information. We shall not 
make use of (10) in this paper.

\subsection{The Gallavotti-Cohen functional}
Note that
\[
\GC_t(\alpha)=\lim_{s\rightarrow\infty}\lim_{M\rightarrow\infty}\log
\omega_{s,M}\left(\e^{-\alpha t\Sigma_{M}^t}\right).
\]
Since
\[
\omega_{s,M}\left(\e^{-\alpha t \Sigma_{M}^t}\right)
=\frac{\det (1 + \e^{k_{-s,M}/2}\e^{\alpha(k_{t,M}- k_{M})}\e^{k_{-s,M}/2})}{\det(1 +\e^{k})},
\]
the arguments of the last two sections yield
\[
\lim_{M\rightarrow \infty}\log \omega_{s,M}\left(\e^{-\alpha t \Sigma_{M}^t}\right)
=-t\int_0^\alpha\d\gamma\int_0^1\d u\,\tr\left(\left(1+\e^{-\gamma(k_{t(1-u)}-k_{-tu})}
\e^{-k_{-(s+tu)}}\right)^{-1}\i[k,h]\right).
\]
The operator $k$ is bounded, commutes with $h_0$ and $\Ran(k)\subset\fh_\ac(h_0)$. It follows that 
\begin{equation}
\slim_{s\rightarrow\pm\infty}k_s
=\slim_{s\rightarrow\pm\infty}\e^{\i sh}\e^{-\i sh_0}k\e^{\i sh_0}\e^{-\i sh}
=w_{\pm}kw_\pm^\ast=k_\pm,
\label{k-lim}
\end{equation}
and the dominated convergence theorem yields 
\beq
\GC_t(\alpha)=t\int_0^\alpha\d\gamma\int_0^1\d u\,\tr\left({\cal K}_{{\rm GC},t}(\gamma,u)
\i[k,h]\right),
\label{der-form-gc}
\eeq
where 
\beq
{\cal K}_{{\rm GC},t}(\alpha, u)=-\left(1+\e^{-\alpha(k_{t(1-u)}-k_{-tu})}\e^{-k_-}\right)^{-1}.
\label{kgc}
\eeq

\subsection{Main results}

{\bf Proof of Theorem \ref{mainthm}.} It follows from \eqref{kes}, \eqref{kgc} and \eqref{k-lim} that 
for $u\in ]0,1[$,
\[
{\cal K}_+(\alpha)=\slim_{t\rightarrow\infty}{\cal K}_{{\rm ES},t}(\alpha,u)
=\slim_{t\rightarrow\infty}{\cal K}_{{\rm GC},t}(\alpha,u)
=-\left(1+\e^{-\alpha (k_+-k_-)}\e^{-k_-}\right)^{-1}.
\]
Hence, by dominated convergence,
\[
e_+(\alpha)=\lim_{t\rightarrow\infty}\frac1t\ES_t(\alpha)
=\lim_{t\rightarrow\infty}\frac1t\GC_t(\alpha)
=\int_0^\alpha\tr({\cal K}_+(\gamma)\i[k, h])\d\gamma.
\]
It follows from \eqref{smatrix} and Theorem 4.1 in \cite{AJPP2} that 
\begin{align*}
\tr&\left({\cal K}_+(\gamma)\i[k, h]\right)
=-\tr\left(\left(1+\e^{-\gamma(s^\ast ks-k)}\e^{-k}\right)^{-1}w_-^\ast\i[k,h]w_-\right)\\[3mm]
&=\int_{\cal E}\tr\left(\left(1+
\e^{-\gamma(s^\ast(E)k_0(E)s(E)-k_0(E))}\e^{-k_0(E)}\right)^{-1}
\left(s^\ast(E)k_0(E)s(E)-k_0(E)\right)\right)\frac{\d E}{2\pi}.
\end{align*}
Theorem 4.1 in \cite{AJPP2} is quite general and in the special case considered here the above 
formula can be also checked by an explicit computation \cite{La}. By Lemma \ref{hiai-petz}, 
\begin{align*}
\tr&\left(\left(1 +\e^{-\gamma(s^\ast(E)k_0(E)s(E)-k_0(E))}
\e^{-k_0(E)}\right)^{-1}\left(s^\ast(E)k_0(E)s(E)-k_0(E)\right)\right)\\[3mm]
&=\frac{\d\ }{\d\gamma}\tr\log \left(1+\e^{k_0(E)/2}
\e^{\gamma (s^\ast(E)k_0(E)s(E)-k_0(E))}\e^{k_0(E)/2}\right),
\end{align*}
and so, Fubini's theorem yields
\[
\begin{split}
e_+(\alpha)&=\int_0^\alpha \tr ({\cal K}_{+}(\gamma)\i [k, h])\d \gamma\\[3mm]
&=\int_{\cal E}\left[\int_0^\alpha\frac{\d}{\d\gamma}\tr\log \left(1+\e^{k_0(E)/2}
\e^{\gamma (s^\ast(E)k_0(E)s(E)-k_0(E))}\e^{k_0(E)/2}\right)\d\gamma\right]\frac{\d E}{2\pi}
\\[3mm]
&= \int_{\cal E}\log \frac{\det \left(1 + \e^{k_0(E)/2}
\e^{\alpha(s^\ast(E)k_0(E)s(E)-k_0(E))}\e^{k_0(E)/2}\right)}
{\det(1 +\e^{k_0(E)})}\frac{\d E}{2\pi}.
\end{split}
\]
The formula for $e_{p,+}(\alpha)$ is derived in the same way, starting with
\begin{align*}
{\cal K}_{p,+}(\alpha)&=\slim_{t\rightarrow \infty}{\cal K}_{p,t}(\alpha, u)\\[3mm]
&=\frac{1}{2} \e^{-(1-\alpha)k_+/p}\left(1+\left(\e^{(1-\alpha)k_+/p}\e^{2\alpha k_-/p}
\e^{(1-\alpha)k_+/p}\right)^{-p/2}\right)^{-1}\e^{(1-\alpha)k_+/p}  + {\rm h.c.},
\end{align*}
for $p\in]0,\infty[$, and
\[
{\cal K}_{\infty,+}=\slim_{t\rightarrow \infty} {\cal K}_{\infty,t}(\alpha, u)
=\left(1 +\e^{-(1-\alpha)k_+-\alpha k_-}\right)^{-1}.
\]
This, together with the remark preceding Theorem \ref{mainthm}, prove Part (1).

The formulas derived in Part (1) clearly show that $e_{p,+}$ and $e_+$ are real analytic functions 
of $\alpha$ satisfying $e_{p,+}(0)=e_+(0)=0$. As limits of convex functions they are also convex and 
hence have non-negative second derivatives. Thus, these derivatives either vanish identically or have an 
isolated set of real zeros. It follows that  $e_{p,+}$ and $e_+$ are either linear or strictly convex . The
symmetry $e_{p,+}(\alpha)=e_{p,+}(1-\alpha)$, a consequence of Proposition \ref{pain} (4), implies
$e_{p,+}(1)=0$ and hence $e_{p,+}$ is strictly convex if it doesn't vanish identically. Regarding $e_+$,
an explicit calculation (see \cite{La}) shows that if it doesn't vanish identically, then $e_+^{\prime\prime}(\alpha)>0$ for 
all $\alpha\in\rr$.

It follows from \eqref{der-form-td}, \eqref{der-form-gc} that there exists $\epsilon>0$ such that the 
functions $e_{2,t}(\alpha)/t$, $\ES_t(\alpha)/t$ and $\GC_t(\alpha)/t$ have analytic extensions to the
complex disc $|\alpha|<\epsilon$ and are uniformly bounded on this disc for $t>0$. 
Vitali's convergence theorem, Proposition \ref{therm2} (7) and the fact that 
$\GC_t^\prime(0)=-t\langle\sigma\rangle_+$ imply
$$
e_{p,+}^{\prime}(0)=\lim_{t\rightarrow\infty}\frac1t e_{p,t}^{\prime}(0)
=\lim_{t\rightarrow\infty}\frac1t\ES_t^{\prime}(0)=e_+^\prime(0)
=\lim_{t\rightarrow\infty}\frac1t\GC_t^{\prime}(0)=-\langle\sigma\rangle_+.
$$
Moreover,
\[
e_+^{\prime\prime}(0)=\lim_{t\rightarrow\infty}\frac1t\ES_t^{\prime\prime}(0)
=\lim_{t\rightarrow\infty}\frac1t\GC_t^{\prime\prime}(0),
\qquad
e_{2,+}^{\prime\prime}(0)=\lim_{t\rightarrow\infty}\frac1t e_{2,t}^{\prime\prime}(0),
\]
and since $e_{2,t}^{\prime\prime}(0)=\ES_t^{\prime\prime}(0)$ by Proposition \ref{therm2} (8), 
we derive that  $e_{2,+}^{\prime\prime}(0)=e_+^{\prime\prime}(0)$. 
Using that $\omega_+$ is $\tau^t$-invariant, one easily derives 
\begin{align*}
\frac1t\GC_t^{\prime\prime}(0)
&=\frac{1}{2}\int_{-t}^t\left<(\sigma_s-\langle\sigma\rangle_+)(\sigma-\langle\sigma\rangle_+)\right>_+
\left(1-\frac{|s|}{t}\right)\d s\\[4pt]
&=\frac1t\int_{0}^t\left[\frac12\int_{-s}^s
\left<(\sigma_u-\langle\sigma\rangle_+)(\sigma-\langle\sigma\rangle_+)\right>_+\d u\right]\d s.
\end{align*}
This completes the proof of Parts (2) and (3).

To prove (4), we compare $e_+(1)$ and $e_{\infty,+}(1)$. Since $\det(1+A)=1+\tr(A)+\det(A)$
for any $2\times 2$ matrix $A$ and $\det(K_\alpha(E))=\det(K_{\alpha,\infty}(E))$, we have, taking into
account Remark 2 after Theorem \ref{mainthm},
\[
\det\left(1+K_\alpha(E)\right)-\det\left(1+K_{\alpha,\infty}(E)\right)
=\tr\left(\e^{k_0(E)}\e^{\alpha\Delta(E)}\right)-\tr\left(\e^{k_0(E)+\alpha\Delta(E)}\right),
\]
with $\Delta(E)=s^\ast(E)k_0(E)s(E)-k_0(E)$. By the Golden-Thompson inequality (see Corollary 2.3 
and Exercise 2.8 in \cite{JOPP}), the above difference of traces is strictly positive unless 
\beq
[k_0(E),\Delta(E)]=[k_0(E),s^\ast(E) k_0(E)s(E)]=0.
\label{comeq}
\eeq
One easily verifies that for $E\not=0$ this happens if and only if $s(E)$ is either diagonal or 
off-diagonal. It follows that 
\[
e_+(1)>e_{\infty,+}(1)=0,
\] 
unless $h$ is reflectionless.

To prove Part (5), note that  
\[
\det(1+K_{\alpha,p}(E))-\det(1+K_{\alpha,q}(E))=\tr(K_{\alpha,p}(E))-\tr(K_{\alpha,q}(E)).
\]
The Araki-Lieb-Thirring inequality (\cite{Ar5, LT}, see also Theorem 2.2. and Exercise 2.8 in
\cite{JOPP}) implies that the function 
$$
]0, \infty]\ni p\mapsto\tr(K_{\alpha,p}(E))=\tr\left(\left[
\e^{(1-\alpha)k_0(E)/p}\e^{2\alpha s(E)k_0(E)s^\ast(E)/p}\e^{(1-\alpha)k_0(E)/p}\right]^{p/2}\right),
$$ 
is strictly decreasing unless \eqref{comeq} holds, in which case
this function is constant. Part (5) follows. \qed

{\bf Proof of Corollary \ref{mainthm-1}.}
Part (1) follows from Theorem \ref{mainthm} and the G\"artner-Ellis theorem (see Appendix A.2 in
\cite{JOPP} or any book on large deviation theory). The analyticity argument used
in the proof of Part (3) of Theorem \ref{mainthm} and Bryc lemma (see \cite{Br} and Appendix A.4. in
\cite{JOPP}) yield Part (2). \qed



\begin{thebibliography}{99999999}

\bibitem[Ar1]{Ar1} Araki, H.: Relative entropy of states of von Neumann algebras. 
Publ. Res. Inst. Math. Sci. Kyoto Univ. {\bf 11}, 809 (1975/76).
  
\bibitem[Ar2]{Ar2} Araki, H.: Relative entropy of states of von Neumann algebras II. 
Publ. Res. Inst. Math. Sci. Kyoto Univ. {\bf 13}, 173 (1977/78).

\bibitem[Ar3]{Ar3} Araki, H.: On the XY-model on two-sided infinite chain.
Publ. Res. Inst. Math. Sci. Kyoto Univ. {\bf 20}, 277 (1984).

\bibitem[Ar4]{Ar4} Araki, H.: Master symmetries of the XY model. 
Commun. Math. Phys. {\bf 132}, 155 (1990).

\bibitem[Ar5]{Ar5} Araki, H.: On an inequality of Lieb and Thirring. 
Lett. Math. Phys. {\bf 19}, 167 (1990).

\bibitem[Av]{Av} Avila, A.: The absolutely continuous spectrum of the almost Mathieu operator. 
Preprint.

\bibitem[AB1]{AB1} Aschbacher, W., Barbaroux, J.-M.: Out of equilibrium correlations in the XY chain.
Lett. Math. Phys. {\bf 77}, 11 (2007).

\bibitem[AB2]{AB2} Araki, H., Barouch, E.: On the dynamics and ergodic properties of the XY-model. 
J. Stat. Phys. {\bf 31}, 327 (1983). 
  
\bibitem[AH]{AH} Araki, H., Ho, T.G.: Asymptotic time evolution of a partitioned infinite two-sided
isotropic XY chain. Proc. Steklov Inst. Math. {\bf 228}, 191 (2000).

\bibitem[AP]{AP} Aschbacher, W., Pillet, C-A.: Non-equilibrium steady states of the XY chain.
J. Stat. Phys. {\bf 112}, 1153 (2003).

\bibitem[AJJP1]{AJPP1}Aschbacher, W., Jak\v si\'c, V., Pautrat, Y., Pillet C.-A.: Topics in
non-equilibrium quantum statistical mechanics. In {\it Open Quantum System III. Recent 
Developments.} S.~Attal, A.~Joye and C.-A.~Pillet editors. 
Lecture Notes in Mathematics, {\bf 1882}, Springer, Berlin,  2006.

\bibitem[AJPP2]{AJPP2} Aschbacher W., Jak\v si\'c V., Pautrat Y., Pillet C.-A.: Transport properties of
quasi-free fermions. J. Math. Phys. {\bf 48}, 032101 (2007).

\bibitem[Bi]{Bi}Billingsley, P.: {\sl Probability and Measure.} Second edition. Wiley, New York, 1986.

\bibitem[BM]{BM} Barouch, E., McCoy, B.M.: Statistical mechanics of the XY model II. Spin-correlation
functions. Phys. Rev. A {\bf 3}, 786 (1971). 

\bibitem[BSZ]{BSZ} Baez, J.C., Segal, I.E., Zhou, Z.: {\em Introduction to Algebraic and Constructive 
Quantum Field Theory.} Princeton University Press, Princeton NJ, 1991.

\bibitem[BR1]{BR1} Bratteli, O., Robinson, D. W.: {\em Operator Algebras and Quantum
Statistical Mechanics 1.} Springer, Berlin, 1987.

\bibitem[BR2]{BR2} Bratteli, O., Robinson, D. W.: {\em Operator Algebras and Quantum Statistical
Mechanics 2.} Springer, Berlin, 1996.

\bibitem[Br]{Br} Bryc, W.: A remark on the connection between the large
deviation principle and the central limit theorem. Stat. Prob. Lett. {\bf 18}, 253 (1993). 

\bibitem[CG]{CG} Cohen, E.G.D., Gallavotti, G.: Dynamical ensembles in nonequilibrium statistical 
mechanics. Phys. Rev. Lett. {\bf 74}, 2694 (1995).

\bibitem[CWWSE]{CWWSE} Carberry D.M., Williams S.R., Wang G.M., Sevick E.M., Evans D.J.:
The Kawasaki identity and the fluctuation theorem.
J. Chem. Phys. {\bf 121}, 8179--8182 (2004). 

\bibitem[DDM]{DDM} Derezi\'nski, J., De Roeck, W., Maes, C.: Fluctuations of quantum currents and
unravelings of master equations. J. Stat. Phys. {\bf 131}, 341 (2008).

\bibitem[HP]{HP} Hiai, F., Petz, D.: The Golden-Thompson trace inequality is complemented. 
Lin. Alg. Appl. {\bf 181}, 153 (1993).

\bibitem[HR]{HR} Hume, L., Robinson, D.W.: Return to equilibrium in the XY model. 
J. Stat. Phys. {\bf 44}, 829 (1986).

\bibitem[ES]{ES} Evans, D.J.,  Searles, D.J.: Equilibrium microstates which generate second law 
violating steady states. Phys Rev. E {\bf 50}, 1645 (1994).
 
\bibitem[I]{I} Israel, R.: {\em Convexity in the Theory of Lattice Gases.} Princeton Series in Physics,
Princeton Univ. Press, Princeton, NJ, 1979.
 
\bibitem[J]{J} Jak\v si\'c, V.: Topics in spectral theory. In {\it Open Quantum Systems I. The
Hamiltonian Approach.} S.~Attal, A.~Joye and C.-A.~Pillet editors. Lecture Notes in Mathematics
{\bf 1880}, Springer, Berlin,  2006.

\bibitem[JKP]{JKP} Jak\v si\'c, V., Kritchevski, E., Pillet, C.-A.: Mathematical theory of the
Wigner-Weisskopf atom. In {\em Large Coulomb Systems.} J.~Derezi\'nski and H.~Siedentop editors.
Lecture Notes in Physics {\bf 695}, Springer, Berlin, 2006.

\bibitem[JOP]{JOP}Jak\v si\'c, V., Ogata, Y., Pillet, C.-A.: Entropic fluctuations in statistical
mechanics II. Quantum dynamical systems. In preparation. 

\bibitem[JOPP]{JOPP} Jak\v si\'c, V., Ogata, Y., Pautrat, Y., and Pillet, C.-A.:
\newblock Entropic fluctuations in quantum statistical mechanics--an introduction.
\newblock In {\sl Quantum Theory from Small to Large Scales.}
\newblock J.~Fr\"ohlich, M.~Salmhofer, V.~Mastropietro, W.~De Roeck
and L.F.~Cugliandolo editors.
\newblock Oxford University Press, Oxford, 2012.

\bibitem[JP1]{JP1} Jak\v si\'c V., Pillet, C.-A.: On entropy production in quantum statistical
mechanics. Commun. Math. Phys. {\bf 217}, 285 (2001).

\bibitem[JP2]{JP2} Jak\v si\'c, V., Pillet, C.-A.: Mathematical theory of non-equilibrium
quantum statistical mechanics. J. Stat. Phys. {\bf 108}, 787 (2002).

\bibitem[JP3]{JP3} Jak\v si\'c, V., Pillet, C.-A.: Entropic functionals in quantum statistical mechanics.
To appear in: Proceedings of the XVIIth International Congress of Mathematical Physics, 
Aalborg, Denmark, 2012.

\bibitem[JPR]{JPR} Jak\v si\'c, V., Pillet C.-A, Rey-Bellet, L.: Entropic fluctuations in statistical
mechanics I. Classical dynamical systems. Nonlinearity {\bf 24}, 699 (2011).

\bibitem[JW]{JW} Jordan, P., Wigner, E.: Pauli's equivalence prohibition. 
Z. Physik {\bf 47}, 631 (1928).

\bibitem[Ku]{Ku} Kurchan, J.: A quantum fluctuation theorem.   
Arxiv preprint cond-mat/0007360 (2000).

\bibitem[La]{La} Landon, B.: Master's thesis, McGill University. In preparation. 

\bibitem[LL]{LL} Levitov, L.S.,  Lesovik, G.B.:  Charge distribution in quantum shot noise. 
JETP Lett. {\bf 58}, 230 (1993).

\bibitem[LSM]{LSM} Lieb, E., Schultz, T., Mattis, D.: Two solvable models of an antiferromagnetic 
chain. Ann. Phys. {\bf 16}, 407  (1961).

\bibitem[LT]{LT} Lieb, E.,Thirring, W.: Inequalities for the moments of the eigenvalues of the
Schr\"odinger Hamiltonian and their relation to Sobolev inequalities. In Studies in Mathematical
Physics. E.~Lieb, B.~Simon and A.S.~Wightman editors. Princeton University Press, Princeton,
NJ, 1976.

\bibitem[Ma]{Ma} Matsui, T.: On conservation laws of the XY model. 
Math. Phys. Stud. {\bf 16}, 197 (1993).

\bibitem[Mc]{Mc} McCoy, B.M.: Spin correlation functions of the XY model. 
Phys. Rev {\bf 173}, 531 (1968).

\bibitem[OM]{OM} Ogata, Y., Matsui, T.: Variational principle for non-equilibrium steady states of XX
model. Rev. Math. Phys. {\bf 15}, 905  (2003).

\bibitem[OP]{OP} Ohya, M., Petz, D.: {\em Quantum Entropy and its Use.} Springer, Berlin, 2004.

\bibitem[Re]{Re} Remling, C.: The absolutely continuous spectrum of Jacobi matrices. 
Annals of Math. {\bf 174}, 125 (2011).

\bibitem [Ro]{Ro} de Roeck, W.: Large deviation generating function for currents in the
Pauli-Fierz model. Rev. Math. Phys. {\bf 21}, 549 (2009).

\bibitem[RS]{RS} Reed, M., Simon, B.: {\em Methods  of Modern Mathematical Physics, III.
Scattering Theory}, Academic Press, London, 1978. 

\bibitem[RM]{RM} Rondoni, L., Me\'\myi ja-Monasterio, C.: Fluctuations in non-equlibrium statistical
mechanics: models, mathematical theory, physical mechanisms. Nonlinearity {\bf 20}, 1 (2007).

\bibitem[Ru1]{Ru1} Ruelle, D.: {\em Statistical mechanics. Rigorous result.} 
Benjamin, New York, 1969.

\bibitem[Ru2]{Ru2} Ruelle, D.: Entropy production in quantum spin systems.  
Commun. Math. Phys. {\bf 224}, 3 (2001).
 
\bibitem[S]{S} Simon, B.: {\em The statistical mechanics of lattice gases, I.} Princeton University
Press, Princeton, NJ, 1993.
 
\bibitem[TM]{TM}  Tasaki, S.,  Matsui, T.:  Fluctuation theorem, nonequilibrium steady states and
MacLennan-Zubarev ensembles of a class of large quantum systems. Quantum Prob. White
Noise Anal. {\bf 17}, 100 (2003). 

\bibitem[Te]{Te} Teschl, G.: {\em Jacobi Operators and Completely Integrable
Nonlinear Lattices.} Mathematical Surveys and Monographs {\bf 72}, AMS, Providence 1991.

\bibitem[N]{N} Nenciu, G.: Independent electrons model for open quantum systems: 
Landauer-B\"uttiker formula and strict positivity of the entropy production.  
J. Math. Phys. {\bf 48}, 033302 (2007).

\bibitem[VY]{VY} Volberg, A., Yuditskii, P.: On the inverse scattering problem for Jacobi matrices with
the spectrum on an interval, a finite system of intervals, or a Cantor set of positive length. 
Commun. Math. Phys. {\bf 226}, 567 (2002).

\bibitem[Y]{Y} Yafaev, D.R.: {\em Mathematical Scattering Theory. General Theory.} 
Translated from the Russian by J. R. Schulenberger. Translations of Mathematical Monographs, 105. 
American Mathematical Society, Providence, RI, 1992.
\end{thebibliography}
\end{document}